\documentclass[aps,prd,12pt, groupedaddress,eqsecnum,amssymb,showpacs,
nofootinbib]{revtex4}
\usepackage{graphicx,mathptm,bm,times,epsfig,amsmath}

\newcommand{\bea}{\begin{eqnarray}}
\newcommand{\eea}{\end{eqnarray}}

\begin{document}

\title{Entropy of the Information Retrieved from Black Holes}

\author{Laura Mersini-Houghton}
\email[]{mersini@physics.unc.edu}
\affiliation{Department of Physics and Astronomy, UNC-Chapel Hill,
NC, 27599-3255, USA}

\date{\today}
 
\begin{abstract}

The retrieval of black hole information was recently presented in two interesting proposals in the 'Hawking Radiation' conference: a revised version by G. 't Hooft of a proposal he initially suggested 20 years ago and, a new proposal by S. Hawking. Both proposals address the problem of black hole information loss at the classical level and derive an expression for the scattering matrix. The former uses gravitation back reaction of incoming particles that imprints its information on the outgoing modes. The latter uses supertranslation symmetry of horizons to relate a phase delay of the outgoing wave packet compared to their incoming wave partners. The difficulty in both proposals is that the entropy obtained from them appears to be infinite.

By including quantum effects into the Hawking and 't Hooft's proposals, I show that a subtlety arising from the inescapable measurement process, the Quantum Zeno Effect, not only tames divergences but it actually recovers the correct $1/4$ of the area Bekenstein-Hawking entropy law of black holes. 

\end{abstract}


\maketitle

\section{Introduction} 
\label{sec:intro}

The mystery of black hole information loss was recently addressed by S. Hawking's new proposal \cite{shawkingnew} presented at the Hawking Radiation conference I organized in Sweden, and by G 't Hooft's diagonalized brick wall model \cite{thooft}. Although the two models differ in their approaches, they both treat the black hole information retrieval problem classically. S.Hawking proposes the surface of the black hole to be like a hologram and uses supertranslation symmetry from the BMS group to gain information about the phase delay of a wavepacket scattered from the horizon \cite{shawkingnew}. G. 't Hooft uses the gravitational barrier to postulate an imaginary surface near the horizon, a 'brick wall', from which incoming waves scatter off as if bouncing from an invisible 'hard wall'  \cite{thooft} and the outgoing modes contain the information of near horizon scattering in their phase. 

Both models aim to extrapolate a scattering matrix relating the in and out going waves. While successfully constructing the $S - $matrix elements, both models run into a serious difficulty, the black hole entropy diverges, which implies an infinite number of microstates. The phase delay for the outgoing wavepackets in both proposals depends on momenta and the transverse coordinates. The divergence at the classical level is to be expected since the number of modes being scattered is unbound and their energies are unbound, (becoming worse near the horizon due to blueshifting). These lead to a quadratic divergence in the transverse momenta as well as a logarithmic divergence. 

Achieving an $S$ matrix description to relate the incoming and outgoing wavepackets makes these proposals the most promising solution to a $45$ years old outstanding problem. So, it is reasonable to consider these models seriously and hope that when quantum effects are incorporated in the two proposals, the entropy of the black hole derived from them will recover the celebrated Bekenstein-Hawking entropy area law.A remaining alternative to the information loss puzzle is that singularities and event horizons may not exist as considered in \cite{lmhbh}, a question that will soon be settled observationally by the Event Horizon Tracer (EHT).

The main emphasis in both models \cite{shawkingnew, thooft} is on the information encapsulated in the phase of the outgoing wavepackets about the incoming modes scattered near the horizon.

Here, I am interested in the quantum aspects of the problem, specifically an additional subtlety related to quantum measurement. When the scattering is treated at the quantum level, then the set of incoming and outgoing modes are inevitably subjected to measurement processes by the black holes, occuring during their scattering off or near the horizon. Let us consider the black hole with mass $M$ as a classical measuring device (a detector), 'watching over', that is, performing random measurements onto the incoming quantum modes to determine their state. The frequency of measurement depends on the frequency of the wavepacket. If we ignore the effect of measurement then the evolution of the scattered modes is given in \cite{shawkingnew, thooft}.

For the sake of illustrating the effect of measurement I will consider the 't Hooft model \cite{thooft} as a working example, since the details of the S. Hawking's proposal \cite{shawkingnew} presented at the Hawking Radiation' conference are not yet available to the public. However without loss of generality, the effect of measurement on the entropy and probability discussed here in the context of \cite{thooft} would apply in the same manner to model for information retrieval of \cite{shawkingnew}.  

\section{Entropy of the Information Retrieved and Quantum Zeno Effect}

\subsection{G. 't Hooft's Brick Wall model}

Let's briefly review the key findings and notation of \cite{thooft} needed for the next section. The reader is referred to the original papers \cite{thooft} for the model and for the steps of derivation of formulas used below. 

The transverse coordinates ($\theta , \phi$) are denoted by $\tilde{x}$ and the longitudinal ones by $z^{\pm} =\frac{1}{\sqrt{2}} (z \pm t)$. The flat metric is written in terms of $u^{\mu} = (\tilde{x}, z^{\pm} )$. The author of \cite{thooft} showed that if a particle $A$, for example with a momentum $p^{\mu} = (0,0,p^{-}, 0)$ travels through a reference point $(0,0,0,0)$ it will induce a shift along the $p^{-}$ direction to some test particle $B$ sitting at $(\tilde{x},0,0)$ by displacing it to a new position $(\tilde{x}, - 4 G p^{-} Log[\frac{\tilde{x}}{C}],0)$, where $G$ is Newton's constant and $C$ a constant. Note that the shift depends on the transverse coordinates. One can think of this effect as the shock wave of the fast moving particle $A$ being produced near the horizon, hits particle $B$ and displaces it by the above amount.

In \cite{thooft} a scattering matrix is derived by using the shift of the outgoing modes that a particle going in to the black hole induces, then generalizing it to a distribution of in and outgoing particles, from which the $|in>$  and $|out>$ states of the black hole (with the added in/out particles), are constructed

\begin{equation}
< out|S|in > = < out_{0}| S |in_{0} > e^{4 i G \int d^2 \tilde{x} Log[(\tilde{x'} - \tilde{x})/C] p_{out}(\tilde{x'})^{+} p_{in}(\tilde{x})^{-}}
\label{scattering} 
 \end{equation}

In this expression \ref{scattering}, $p_{in}^{-}, p_{out}^{+} (\tilde{x})$ describe the total momentum distribution on the horizon of in and out going particles.

Using Rindler coordinates $(\rho, \tau)$, after diagonalizing for the scattered modes \cite{thooft}, a 'hard wall' boundary condition is imposed at $\rho =\rho_{0}$ where the mode bounces. This boundary condition gives a relation between the in and the outgoing mode that bounced off the 'wall' at location $\rho_{0}$ near the horizon.The relation between the diagonalized Fourier transformed in and out going modes $\psi_{i}$ becomes

\begin{equation}
\psi_{i}(k)^{out} = A_i(k)^{\*} e^{ - i k Log(\lambda)} \psi_{i}(k)^{in}
\label{mode}
\end{equation}

where $k$ is the Fourier parameter, $A_i(k)$ the coefficient, and $\lambda = \frac{8\pi G}{l^2}$, where $l$ is the transverse momenta. The exchange of the transverse momentum $l$ is assumed to be small and ignored.
The system is now placed in a box with size $\rho_{1}= 1/2 Log[\Lambda] \simeq Log[M^2]$ and the boundary conditions in the box provide the energy levels $k_n$ by

\begin{equation}
\pi n = k_{n} (\rho_{1} - \rho_{0} ) \simeq k_{n} [ Log(\Lambda) - 1/2 Log(\lambda) -1/2 Log(k) +1)
\end{equation}

From here, assuming the size of the box is large and ignoring the transverse momentum, the density matrix, or partition function in this classical picture, becomes

\begin{equation}
Z = e^{-\beta F}= \Sigma_{states} e^{-\beta E_{n}} \simeq \Sigma_{k_{n}} e^{-\beta k_{n}} = \frac{1}{\pi\beta} [ 2 Log(\Lambda) + Log(\beta) + \gamma - Log(\lambda)]
\label{partitionfunction}
\end{equation}

Here $\beta = 8 \pi M$ is the Hawking inverse temperature, $\gamma$ is the Euler Gamma coefficient, $Log[\Lambda]$ is roughly the size of the box taken to recover Hawking radiation at infinity $\Lambda \simeq M^2$, and the reason for the sum over states being $\simeq$ rather than equal to the sum over the 'energy levels $k_n$ is due to the transverse momentum not being included in the above sum.

\subsection{Black Hole Entropy in the context of the Quantum Zeno Effect}

Each mode with an energy level $k_n$ will have a $(2j+1)$ degeneracy due to angular momentum $j$ that needs to be summed. So, the sum over states would normally include a $\Sigma_{j} j(j+1)$. Equivalently in the continuum limit we need to integrate with respect to the transverse momenta $l$. Then Eq. \ref{partitionfunction} contains a double integral, $\int l dl $ and the integral over $k_n$ in . The problem is that the integration over the transverse momentum $l$ makes the expression quadratically divergent, besides the logarithmic divergence coming from the $Log(\lambda)$ already present in Eq.\ref{partitionfunction}.

An ingoing particle will continue its typical infalling trajectory into a black hole, while the 'bounce at the horizon or the hard wall' refers only to the information carried by the outgoing mode, as emphasized in \cite{thooft}. Therefore both, the 'hologram' and 'brick wall' models allow us to continue alternate between a description of Hawking radiation as vacuum pair creation of entangled particles in the exterior of the black hole, or the alternative description of Hawking radiation as a tunneling process \cite{wilczek}, which is useful for illustrating the quantum effects here. Seen as a tunneling process, black hole radiance is obtained either by outgoing positive energy particles tunneling outside the gravitational barrier, or as a pair creation outside the black hole with the negative energy particle in the pair tunneling into the black hole. We are dealing at the quantum level with unstable quantum systems since particles can tunnel in or out the gravitational potential. 

We would now like to show that with the measurement process the black hole performs on the in and out going modes, high energy modes are trapped while low energy ones decohere. What this means in the tunneling view is: the tunneling lifetime of the high energy particles becomes very long due to their frequent measurement by the black hole. These pairs keep returning to their intial state every time they are monitored. So, quantum entaglement of the high energy pair is preserved, but the particle carrying the entanglement information cannot 'tunnel out' to contribute to Hawking radiation.

To proceed, let us start with the basics. Assume we start with a coherent incoming wavepacket. The scattering matrix element connecting the in and out going states and its 'survival' probability are related up to a factor

\begin{equation}
P_{in,out} \simeq  |< out | S^{+} |in >|^2
\label{probability}
\end{equation}

Since we are interested on the quantum aspect of scattering, we have to take into account the monitoring of the incoming and outgoing modes by the black hole near the bounce point. Let us take a black hole with mass $M$ as a classical detector monitoring the quantum particles. The random measurement the black hole performs on these modes is inevitable and its frequency depends on the frequency of the mode as explained above. The measurement gives rise to the quantum Zeno effect (QZE), a fact well known in literature \cite{zeno}, straightforward to derive, and confirmed experimentally for a variety of quantum systems , (for confining tunneling potentials similar to the one here see \cite{zenotunnel}, and other types of potentials see \cite{zenotest} and references therein). 

QZE leads to three regimes, short time (frequent) measurement, intermediate time and long time measurement. The regime is determined by the frequency of measurements compared to the Zeno defined below. For example, modes with inverse frequencies that are shorter than the Zeno time fall under 'short time measurement' behaviour, but modes with inverse frequencies that are much longer than the Zeno time fall under the 'long time measurement' regime, and so on.  A short time expansion of the probability with the QZE included, leads to a modification of the exponential tunneling probability by a quadratic behaviour in time

\begin{equation}
P(t) \simeq  1 - \frac{\tau^2}{\tau_{z}^{2} } \simeq e^{-\frac{\tau^2}{\tau_{z}^{2} }}
\label{quadratic}
\end{equation}

Here $\tau_z$ is the Zeno time. The Fourier transform of \ref{quadratic} gives the characteristic function, i.e the probability weights for each mode in momentum space.
 Due to the relation of the probability to scattering given by Eq.\ref{probability}, the scattering matrix in Eq.\ref{scattering} also needs to be modified by multiplying it with the square root of Eq.\ref{weights}, for the QZE to be taken into account.

The intermediate time regime allows for the typical exponential tunneling decay and the long term regime modifies the exponential with a power law tail.

The regime is determined by the Zeno time, for example the short time measurements are the ones performed during time intervals less the $\tau_{Z}$ . If we denote by $H_{int}$ the interaction Hamiltonian of the 'detector' with the quantum particles, in our case the interaction of the black hole 'hard wall' with the in and out going particles, then the Zeno time can be calculated from: $\tau_{z}^{-2} = < in| H_{int}^2|in >$. 

Both models we are interested in here \cite{thooft, shawkingnew} consider the scattering surface from which the incoming modes bounce, as a hard wall located at $R \simeq 2M$ or near $\rho =\rho_0$. Then the interaction of the black hole 'hard wall' with these modes can well be approximated by the hard sphere potential. It follows that in the hard sphere approximation the Zeno time becomes $\tau_{z} \simeq M $. Then the characteristic function obtained from Eq. \ref{quadratic} results in probability weights for each mode $(k, l)$, that is modified by the QZE quadratic factor given below

\begin{equation}
p(k, l) \simeq M^2 e^{-M^2(l^2 + k^{2})}
\label{weights}
\end{equation}

with $l$ the transverse momenta and $k$ the longitudinal momenta.
The frequency of measurement is determined by the frequency of the particles $\omega$, in the sense that their monitoring by the black hole occurs during a time $\tau \simeq \frac{1}{\omega}$. For massless particles $\omega^2 = k^2 + l^2 $ and since $\tau_{z} \simeq M$ then a short time regime for the massless modes is valid for those modes with $\omega >> M^{-1}$.

Incorporating the QZE modification to probability of (Eq. \ref{weights}) that arises from the black hole measurement, in the density matrix of Eq.\ref{partitionfunction}, is now straightforward, as shown next. 

 Our goal is to estimate the entropy by including the effect of measurement in the density matrix, and summing over all states including the transverse momenta $l$ in the continuum limit. It follows that \ref{partitionfunction} now reads

\begin{equation}
Z = e^{-\beta F}= \Sigma_{states} e^{-\beta E_{n}} p(k, l) = \int dk \int_{0}^{\infty} l dl  e^{-\beta k} M^2 e^{-M^2(l^2 + k^{2})}\left[ Log(\Lambda) -\frac{1}{2}Log(\lambda) - \frac{1}{2} Log(k ) \right]
\label{qzepartition}
\end{equation}

which yields the following expression for the density matrix with $\beta = 8 \pi M$ the inverse Hawking temperature.

\begin{align}
		Z  &= e^{-\beta F} = \\ 
\notag & \hspace{0.001in} = \left(\frac{1}{4 \beta M^2}  e^{\frac{\beta^2}{4 M^2}} \sqrt{\pi } \left[M^2 \sqrt{\frac{\beta^2}{M^2}} \text{erf}\left(\frac{1}{2} \sqrt{\frac{\beta^2}{M^2}}\right) \left(\ln{M^2} \right) +  \right. \right.  \\ 
\notag & \hspace{0.001in} \left.\left. \beta \left(-M  \gamma   \text{erf}\left(\frac{\beta}{2 M}\right) + \sqrt{M^2}\left( \gamma - 2\ln{M^2} + \ln{4 M^2}\right)\right)\right] + \right. \\ 
\notag & \hspace{0.001in} \left.  \sqrt{\pi}\sqrt{M^2} \beta e^{\frac{\beta^2}{4 M^2}}  {}^{}_{1}\text{F}_{1}^{(1,0,0)}\left(0,\frac{1}{2},-\frac{\beta^2}{4 M^2}\right)+\beta^2 {}^{}_{1}\text{F}_{1}^{(1,0,0)}\left(1;\frac{3}{2};\frac{\beta^2}{4 M^2}\right)\right). 
\label{partitionqze}
\end{align}

Equipped with the result in Eq.2.9, we proceed to estimate the entropy $S$ from it
 
\begin{eqnarray}
S = \beta U - V ,  \,  
& & U = \frac{\partial V}{\partial b}
\label{entro}
\end{eqnarray}

where $V= \beta F$, $ {\it _1 F_{1}[a,b,x] }$ is the hypergeometric function and ${\it erf[x]}$ the error function. The result for the entropy, when the QZE measurement effect is taken into account in the \cite{thooft,shawkingnew} models, is shown in Fig.1.

\begin{figure}[!htbp]
\begin{center}
\raggedleft \centerline{\epsfxsize=3.9in \epsfbox{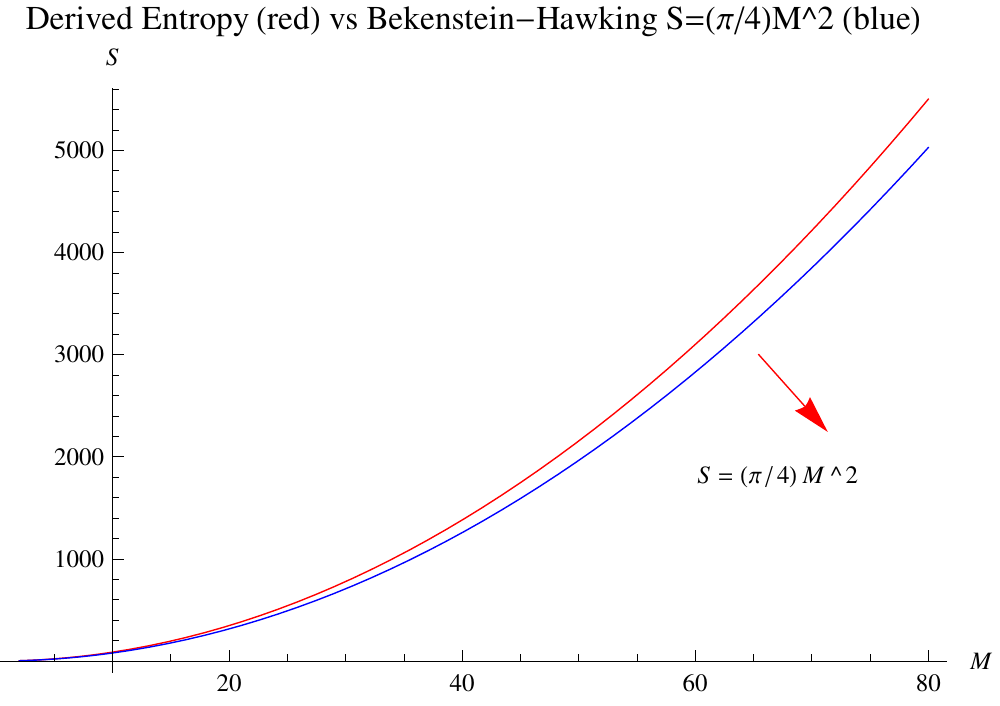}}
\caption{The red color is the derived entropy of the 'Brick Wall' model of \cite{thooft} with the Quantum Zeno Effect included. The blue color plot is the Bekenstein Hawking entropy law of black holes shown for comparison. } 
\label{fig:fig1}
\end{center}
\end{figure}

\section{Remarks}

Although easy to obtain from Eq. 2.9 and \ref{entro}, the entropy expression is algebraically tedious, so instead we show the plot of $S$  in Fig.1 using Mathematica, in order to compare it to the Bekenstein-Hawking entropy area law, plotted in the same figure. It can be seen that the black hole entropy is quite close to the Bekenstein-Hawking entropy. The reason for the resemblance is that the high momenta modes are trapped by the black hole monitoring thus their contribution to the entropy is highly dampened. 

Physically both models collect the information about the vacuum pairs entanglement in the Hawking radiation. With the QZE included this process is modified. The high energy pairs are 'stuck' in their initial quantum entangled state, their frequent measurement halts decoherence protects their entanglement, so information can not be carried away by an outgoing particle and these pairs not contribute significantly to Hawking radiation. As an aside, the QZE seems to provide a natural mechanism for addressing the transplanckian problem of black holes. On the other hand, the lower energy pairs with wavelengths of order $M$ or longer, undergo a different dynamics: their initial pair entanglement is destroyed, information about their infalling partner is carried away by the outgoing partner in the pair by the mechanisms proposed in \cite{thooft, shawkingnew}. Thus modes with wavelengths of order $M$ or longer provide the dominant contribution to the entropy and Hawking radiation The loss of entanglement in the lower energy pairs is due to their infrequent monitoring by the black hole which triggers decoherence. 

The overall result is a mixed state of mainly lower energy particles of wavelengths of order $M$ or so,  seen as thermal radiation away from the black hole, but which transferred their near the horizon information onto Hawking radiation. Black hole monitoring plays an insignificant effect on these wavepackets due to the fact that their measurement is not frequent, occuring at times of order or larger than $M$. These modes follow the nearly normal scattering and evolution described in \cite{thoogt, shawkingnew}, with the only consequence being a 'memory loss'  of their initial entanglement as the result of the measurement process. Their contribution provides the dominant channel for Hawking radiation. 

We should emphasize that if we start say with coherent incoming wavepackets, it is not surprising that due to QZE we end up with an incoherent contribution added to the coherent states, another way of stating that the outgoing wavepackets are in a mixed state while unitarity for the whole system of quantum particle pairs and the black hole is not violated as is clear from the derivation here. Information about the whole system is not lost but the finite number of microstates, Fig.1, is due to the QZE of measurement.






\medskip



\end{document}